# Modelling thermoelectric performance in nanoporous nanocrystalline silicon

Laura de Sousa Oliveira*[1], Vassillios Vargiamidis*, and Neophytos Neophytou*

*Abstract*— **Introducing hierarchical disorder from multiple defects into materials through nanostructuring is one of the most promising directions to achieve extremely low thermal conductivities and thus improve thermoelectric performance. The success of nanostructuring relies on charge carriers having shorter mean-free-paths than phonons so that the latter can be selectively scattered. Nevertheless, introducing disorder into a material often comes at the expense of scattering charge carriers as well as phonons. In order to determine the tradeoff between the degradation of the lattice thermal conductivity and of the power factor due to this, we perform a theoretical investigation of both phonon and electron transport in nanocrystalline, nanoporous Si geometries. We use molecular dynamics for phonon transport calculations and the non-equilibrium Green's function method for electronic transport. We report on the engineering tradeoff that the porosity (number of pores and their in-between distance) has on the overall thermoelectric performance for the material optimization. We indeed find that the reduction in thermal conductivity is stronger compared to the reduction in the power factor, for the low porosities considered in this study (up to 5 %), and that the *ZT* figure of merit can experience a large increase, especially when grain boundaries are included, compared to just nanoporosity.**

## I. INTRODUCTION

New generation thermoelectric (TE) devices with much higher efficiencies and lower prices could be the key for waste heat recovery — 60% of the world's energy is lost as heat [1] — and usher in a new era of self-powering devices. Thermoelectric performance is quantified by the dimensionless figure of merit $ZT = \sigma S^2 T/(\kappa_e + \kappa_l)$, where $\sigma$ is the electrical conductivity, $S$ is the Seebeck coefficient, $T$ is the operating temperature, $\kappa_e$ is the electronic thermal conductivity, and $\kappa_l$ is the lattice thermal conductivity. To date, nanostructuring is the most promising way to decouple thermal and electrical transport, due to phonons and electrons having different mean-free-paths. Most nanostructuring research has focused on increasing thermoelectric performance by decreasing the lattice thermal conductivity of materials [2, 3]. Hierarchical nanostructuring, where defects of various length scales are

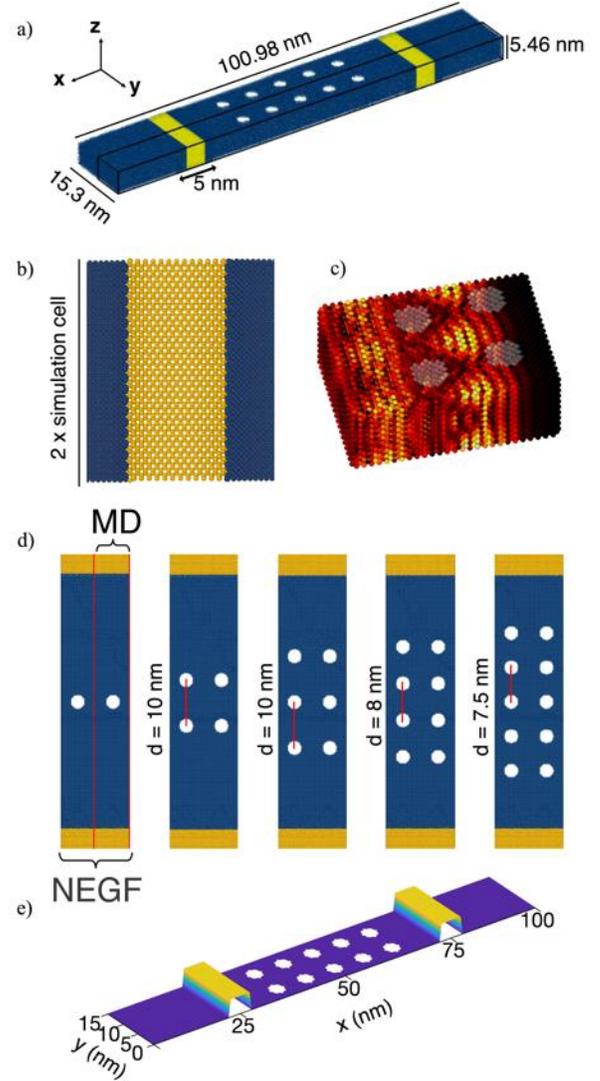

Fig. 1. (a) MD geometry. The rectangular box encasing half of the geometry is the actual MD simulation cell. A single periodic image is shown across the boundary to illustrate how the MD geometry compares to the NEGF channel geometries. (b) Side view of the grain boundaries for two contiguous simulation cells (*i.e.*, across the periodic boundaries). (c) Computer generated graphic illustrating heat flow near the pores. (d) Schematic of the distance between pores for the porous geometries for both electron and phonon calculations. (e) Example of NEGF geometry. Two-dimensional channel with embedded SL barriers and a 5 × 2 array of pores. The SL barriers have a height, $V_{SL}$, of 0.05 eV and a width, $L_B$, of 5 nm.

Paper submitted for review on Feb 1, 2018.
This work has received funding from the European Research Council (ERC) under the European Union's Horizon 2020 Research and Innovation Programme (Grant Agreement No. 678763).

*School of Engineering, University of Warwick, Coventry, CV4 7AL, UK
  [1]L.De-Sousa-Oliveira@warwick.ac.uk,

introduced, has proven particularly effective in lowering the lattice thermal conductivity, $\kappa_l$, due to scattering phonons at different wavelengths [1]. However, lowering $\kappa_l$ is often done at the expense of scattering charge carriers as well, thus reducing the power factor (PF), $\sigma S^2$. Therefore, the importance of retaining high PFs in nanostructured TE materials has also recently come into focus.

The pervasiveness of silicon (Si), not only in its abundance as a natural resource, but in its widespread use in electronics and the corresponding availability of scientific and technological knowledge, make it a desirable material for heat management applications, not only for ICs, but for heat harvesting applications, such as thermoelectrics, as well. A high bulk thermal conductivity, > 100 W m₋₁ K₋₁ [4], has limited the use of silicon in TE devices, despite the moderately high PF — with an upper bound of 6.3 mW m₋₁ K₋₂ at 350 K [5]. However, research has shown a two orders of magnitude reduction in the thermal conductivity of nanostructured Si (e.g. nanowires [6-9] and thin films [10, 11]), as well as Si-based alloys and superlattices [12, 13], compared to bulk Si. As the thermal conductivity has been reduced to nearly the amorphous limit for silicon (i.e., $\kappa < 2$ W m₋₁ K₋₁), $ZT$ values were raised from $ZT_{bulk}$ ~0.01 to ~0.6 [7, 14]. For comparison, good TE materials have $ZT$~1 (10% of Carnot efficiency), but $ZT$~3 is needed for large scale TE applications.

As far as PFs are concerned, Lorenzi et al. [15] have recently measured a surprising very high 22 W m₋₁ K₋₁ PF for heavily boron (B) doped nanocrystalline silicon (n-Si) films of ~30 nm grain sizes embedded with nanovoids. Similar research performed for B doped n-Si without nanovoids yielded also a very high 15 W m₋₁ K₋₁ [16]. Bennett et al. [17] saw a 70% increase in the PF of Si nanolayers when dislocation loops were introduced. The large PF observed in these studies was attributed partially to the combined effect of high doping and dopant segregation on the electrical conductivity ($\sigma$), and carrier energy filtering by means of potential barriers at the grain boundaries on the Seebeck ($S$) coefficient [18]. In short, ways to improve the PF in addition to lowering $\kappa_l$ are reported, which could result in Si-based nanostructured materials becoming viable candidates for thermoelectric applications.

Due to the complexity in accurately simulating the electronic transport at the nanoscale, theoretical works usually focus on one type of nanostructured feature at a time, i.e., only superlattices (SLs) [19, 20], only boundaries [21], only nanoinclusions [22-24], etc. In this work we explore the effect of porosity on the TE coefficients (both conductivity and PF) of materials with SL barriers, i.e., hierarchically disordered materials. We aim to determine the effect of porosity on thermoelectric performance, taking into account its influence on degrading both phonon and electron transport. We begin with what studies by us and others have shown are the optimal conditions for the PF in nanocrystalline materials (i.e., degenerate Fermi levels, and carrier filtering at potential barriers) [25, 26]. To estimate the thermoelectric performance (i.e., compute the figure of merit $ZT$), we combine both electronic and phononic transport calculations. More specifically, we employ equilibrium molecular dynamics (MD) for phonon transport calculations in (3D) geometries (see Fig. 1), and the nonequilibrium Green's function (NEGF) method to calculate the electron transport properties of the equivalent planar 2D nanostructures when both SL type boundaries and/or pores are present (we drop the invariant depth $z$-direction in NEGF for computational efficiency reasons — with justifications that are discussed below). The paper is organized as follows: In Sec. II we describe the computational approach and geometries investigated. In Sec. III we present and discuss our results, and in Sec. IV we summarize the main results and conclusions.

## II. Computational approach

The lattice thermal conductivity calculations are performed within equilibrium MD, with the widely used Green–Kubo [27, 28] method. The lattice thermal conductivity, $\kappa_l$, of a material is related to fluctuations in its heat-current, $J$, through the Green–Kubo formalism, which can be written as:

$$\kappa_{xx} = \frac{V}{k_B T^2} \int_0^\infty < J_{xx}(t) J_{xx}(t + \tau) > d\tau, \qquad (1)$$

where $k_B$ is Boltzmann's constant, $T$ the temperature and $V$ the volume of the simulated region. In (1), the subscript $xx$ denotes the directional thermal conductivity along the $x$-axis, and $< J_{xx}(t) J_{xx}(t + \tau) >$ is the non-normalized heat-current autocorrelation function (HCACF) in the same direction. In this study, transport is investigated along the channels, i.e., in the $x$-direction as shown in Fig. 1a. The HCACF can be computed as the inverse Fourier transform of the same transform of the heat-current (as a function of simulation time) multiplied by its complex conjugate, or numerically as:

$$< J(t) J(t + \tau) > \equiv \sum_{n=0}^{N-m} \frac{J_n J_{n+m}}{N - m}, \qquad (2)$$

where $J_n$ is the value of $J$ at the $n$th time step, and $J_{n+m}$ is $J$ at the ($n+m$)th time step, for n = 0, 1, 2, …, N and m = 0, 1, 2, …., M. Here, N and M are the maximum number of steps in the simulation and in the HCACF, respectively. The systems are 184×14×10 supercells of Si, for a diamond cubic 8 atom unit-cell, which corresponds to a simulation cell with dimensions of 100.98×7.65×5.46 nm₃, as shown in Fig. 1a. The $x$-axis corresponds to the [1 0 0] direction. Periodic boundary conditions are used in all directions. Twelve geometries are considered in total: a pristine system, a superlattice (SL) with two barriers and no pores, 5 geometries with only pores, and 5 SLs with pores. The geometries used for the phonon calculations closely correspond to the ones used for the electronic calculations, with 1.5 nm radius pores, and 5 nm grain boundaries. As a first order investigation, the latter are introduced by simply rotating the silicon 45o around the [0 0 1] direction and are placed so as to be spaced ~50 nm apart (see Fig 1e). As illustrated in Fig. 1a, the MD geometry varies between 1–5 pores by incrementing a single pore. We employ periodic boundary conditions to capture the effect of an infinite length material.

It is important to note that the pristine system has been converged for size, i.e., we have verified that changing the channel dimensions does not yield statistically different thermal conductivity results from those obtained for the present channel dimensions. The atomic forces of silicon are modeled with the Stillinger–Weber [29] potential, using a set of parameters optimized for thermal transport in silicon based on force matching to density functional theory calculations in the generalized gradient approximation (GGA) [30]. The original Stillinger–Weber potential has been widely used to model heat transfer in silicon [31] and has successfully been used to describe elastic constants and thermal expansion coefficients, offering a reasonable match for phonon dispersion relations [32, 33], especially for acoustic phonons. However, it is known to overestimate thermal conductivity [34]. The adjusted potential corrects this issue largely by improving the accuracy of the dispersion relation, especially for the optical modes.

Error is introduced into the calculation of thermal conductivity due to insufficient averaging of the heat-flux, a constraint imposed by the computational expense of MD and the nature of the Green–Kubo approach. For this reason, a compromise has to be made between an earlier HCACF cut-off (instead of truly integrating it to infinity, as in Eq. (1)) that reduces the error in the thermal conductivity, but could neglect slower relaxation processes, or vice-versa. A thorough discussion on the HCACF error can be found in [35]. In this work, a cut-off of 500 ps has been selected for only the pristine system, where slow relaxation processes play a role on the thermal conductivity, and a smaller 150 ps cut-off was selected for all other geometries, given that the HCACF was already converged at this point. Results from sets of 10–15 calculations are averaged for each of the twelve geometries to help mitigate the error.

The calculations are performed with the LAMMPS [36] package. After relaxing the atomic structure and lattice parameters using a conjugate gradient algorithm, each system is given a thermal energy equivalent to 300 K (room temperature), also allowing for thermal expansion, by running it for 125 ps, at a 0.5 fs time step, in the isothermal–isobaric (NPT) ensemble. The systems are then equilibrated in the microcanonical (NVE) ensemble for another 125 ps. Finally, a NVE simulation is run for another 10 ns, with a 2 fs time step, during which time the HCACF is recorded.

For the electronic calculations, we employ the NEGF method. We use a 2D NEGF simulator developed by our group [22]. The NEGF approach takes into account electron-phonon (e-ph) scattering in the self-consistent Born approximation [37]. In this work we consider acoustic (elastic) phonon scattering only. As with the lattice thermal conductivity calculations, twelve geometries are considered in total: a pristine system, a superlattice (SL) with two barriers and no pores, 5 geometries with only pores, and 5 SLs with pores. The number of pores ranges between 2 and 10 for the porous channel geometries. Example geometries are shown in Fig. 1. The channel length is 100 nm, and the width 15 nm. As before, the pores have circular shape with a diameter of 3 nm, but are arranged in $n \times 2$ arrays, where $n$ varies from 1 to 5. The strength of the phonon scattering is adjusted such that the mean free path of electrons is $\lambda = 15$ nm [38]. This adjustment is done by initially simulating a channel with length $L$=15 nm in the ballistic regime, and then increasing the electron-acoustic phonon scattering strength $D_0$ until the channel conductance drops to 50% of its ballistic value. This effectively fixes a mean-free-path of 15 nm for the channel, a value comparable to that of common semiconductors and Si [39-41]. For the SLs, the conduction band, $E_c$, is set to 0 eV and the Fermi-level, $E_F$, is aligned with the height of the SL barriers, i.e., $E_F = V_{SL} = 0.05$ eV, for optimal performance [20, 25, 42], whereas in the pristine channel $E_F = E_c = 0$ eV. In the case of the lower Fermi level, the mean-free-path does not change significantly as expected and only slightly drops to 12 nm — it is well known that under scattering rates proportional to the density of states the mean-free path is constant in energy [16, 43]. We do not consider ionized impurity scattering, which would have resulted from doping. We assume that the $E_F$ is placed at the required position by other means (i.e., modulation doping or gating [44]). However, we discuss the influence of including ionized impurity scattering (IIS) in the Appendix. Essentially, we show that the conductance ratios, $G_{E_F = 0.05eV}/G_{E_F = 0eV}$, are very similar for the electron-phonon limited scattering, and the phonons + IIS scattering cases, although, of course, the absolute conductivity is much lower in the presence of IIS.

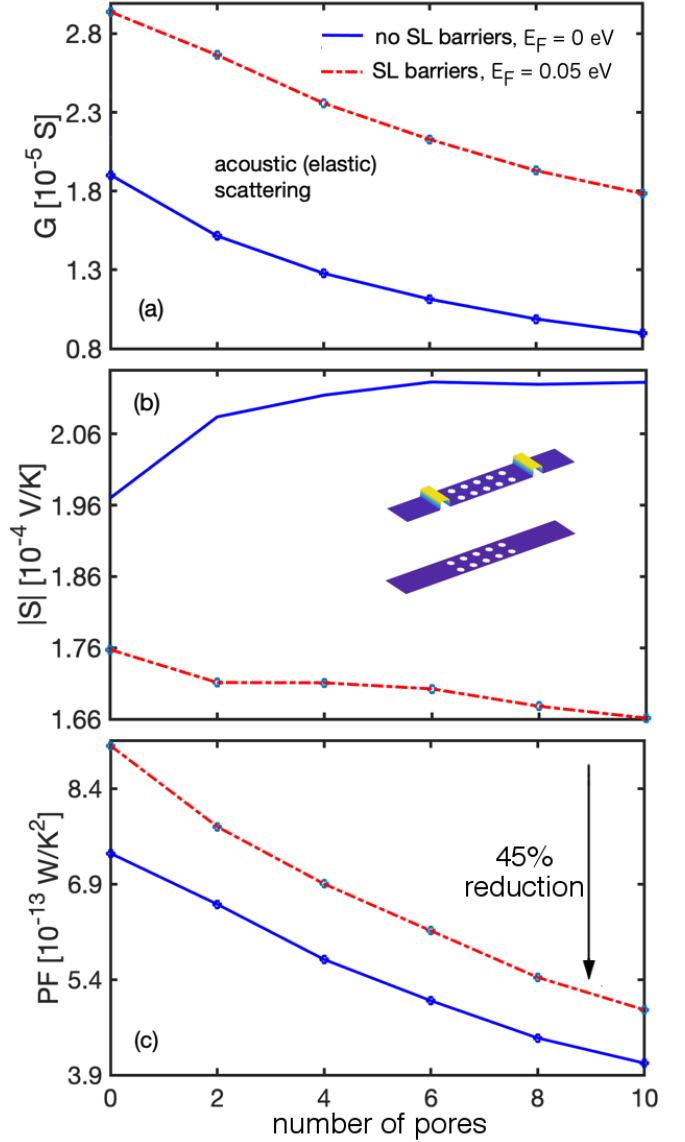

Figure 2. The effect of number of pores N on the thermoelectric coefficients in the presence of AP only (i.e., elastic scattering). (a) The conductance G, (b) the Seebeck coefficient S, (c) the power factor PF, versus the number of pores for a channel with SL barriers (with $E_F = 0.05$eV) and for a pristine channel (with $E_F = 0$ eV).

Additional code details, including channel calibration, can be found elsewhere [22]. In order to calculate the power factor, $GS^2$, where $G$ is the conductance and $S$ the Seebeck coefficient, we use the fact that the Seebeck coefficient is the integral of the average energy of the current flow [25, 42] with respect to $E_F$,

$$S = \frac{1}{qTL}\int_0^L (\langle E(x) \rangle - E_F) dx, \quad (3)$$

where $q$ is the carrier charge ($q = -|e|$ for electrons and $q = +|e|$ for holes) and $\langle E(x) \rangle$ is the average energy of the current flow defined as

$$\langle E(x) \rangle = \frac{\int I(E,x) E dE}{\int I(E,x) dE}, \quad (4)$$

where $I(E,x)$ is the energy and position resolved current.

The electronic and phononic calculations are combined to estimate the overall effects of SLs and/or pores on the thermoelectric figure of merit, $ZT$. Since the electronic calculations are performed for a 2D system (for numerical efficiency), we can therefore calculate only the electrical conductance, $G$, instead of the conductivity, $\sigma$. The lattice thermal conductivity is evaluated in 3D. As a result, we get units that can not be combined to compute $ZT$. Due to having different units for thermal transport, $ZT$ for the pristine system is thus computed using experimental results for $\sigma$, $\kappa_e$ and $\kappa_l$.

We try our best to match the geometries used in MD and NEGF, as well as mean-free-paths for Si in NEGF. Nevertheless, combining the results from MD and NEGF to compute $ZT$ can be problematic. MD simulations are 3D and thus provide us with conductivity, whereas NEGF calculations are 2D and thus provide us with conductance, and equilibrium MD calculations use periodic boundary conditions whereas NEGF simulates a fixed channel. For these reasons, our results for $S$, $G$, $K_e$, and $\kappa_l$ are used to compute the fractional change of the thermoelectric coefficients between the pristine channel and other geometries. This allows us to estimate the change in $ZT$ for the various systems based on our calculations. Thus, we employ literature values for the pristine bulk material TE coefficients, and then scale them according to our calculations for the geometries with features (SL and/or pores) we consider. Thus, ZT is computed as indicated by the following expression:

$$ZT = \frac{\left(\frac{G_{features}}{G_{pristine}}\right)\sigma_{lit} \left[\left(\frac{S_{features}}{S_{pristine}}\right)S_{lit}\right]^2 T}{\left(\frac{K_{e-features}}{K_{e-pristine}}\right)\kappa_{e-lit} + \kappa_{l-lit}\left(\frac{\kappa_{l-features}}{\kappa_{l-pristine}}\right)}, \quad (5)$$

where values from [5] are used where the subscript '*lit*' is indicated in Eq. 5. While these values were obtained at 350 K and our calculations performed at 300 K, we find that they represent the best complete set of parameters. For $\sigma$ we use the value found in [5] (0.588 mΩ-₁ cm-₁) for p-type silicon at 350 K with a $8.1\times10_{19}$ cm-₃ charge carrier concentration. This that best approximates the high PF work of Lorenzi *et al.* as well [15]. For consistency we calculate the electronic thermal conductivity from the above value for $\sigma$ using the Wiedemann–Franz law, yielding $\kappa_{e-literature}$ = 0.5 W m-₁ K-₁. To be consistent, we also use the thermal conductivity on the same

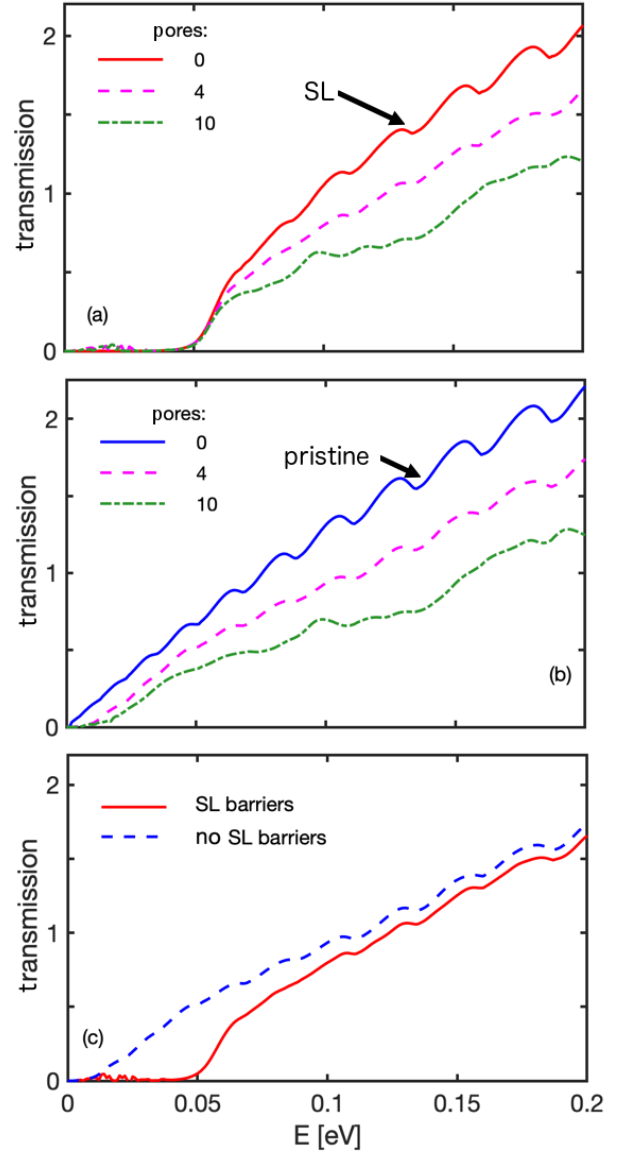

Fig. 3. Transmission versus electron energy for (a) the channel with SL barriers, and (b) the pristine channel, for increasing number of pores. (c) Comparison of the effects of pores on the transmission of the pristine channel and the channel with SL barriers

study, reported as 102 W m-₁ K-₁[5], this leaves us with $\kappa_{l-literature}$ = 101.5 W m-₁K-₁. These values used in conjunction with the value of $S$ obtained in the same study yields a $ZT$ of 0.022, which is on the upper bound of reported values [5], for the pristine system.

Thus, in this comparative approach we take, by calculating the relative change in all the parameters that control $ZT$ (conductivity, Seebeck coefficient, thermal conductivity), and then altering the pristine (experimentally measured) values by those changes, we achieve a reasonable first order estimation of the expected changes in $ZT$. Attempting to compute the absolute values for the $ZT$s, using any combination of simulators, would not be possible: MD overestimates the thermal conductivity to begin with, and the electronic scattering details and their energy dependences (especially at interfaces) would require full atomistic treatment, which is computationally prohibitive within NEGF in the large size structures we consider. Note that for consistency between NEGF and EMD in the geometries calculated, 3D channels within NEGF (of finite size) could have been simulated, or the use of periodic boundary conditions could have been employed, instead of 2D channels. These methods, however, impose an additional and very demanding computational task, especially in the presence of electron-phonon interactions coupling all transverse $y$- and $z$-modes. Since we are interested in the transport coefficient *ratios* between the defected and pristine structures, rather than absolute values, we believe this comparative approach is justified, because obtaining these ratios from 2D NEGF simulations or 3D NEGF simulations, should not differ significantly. 3D simulations would have included the thickness extension (in the $z$-direction as shown in Fig. 1a), however, the structures we consider are geometrically invariant in that direction. Thus, the additional dimension will at first order bring in the calculation of transport only a factor arising from an integration over the solid angle for the charge trajectories and mean-free-paths. Since the same factor would appear in both the pristine and defected channel calculation, it would be canceled out when the ratios of the transport coefficients are computed in Eq. 5. Therefore, as we seek the relative changes due to nanostructuring, details of 2D versus 3D approach 'mixing' will not affect our performance ratios, as each parameter alteration is computed consistently with the nanostructuring itself being the only parameter that affects the changes.

## III. RESULTS

Our simulations aim to explore the effect that changes in thermal conductivity due to porosity have on the thermoelectric efficiency of hierarchical architectures with optimized electronic properties. For this reason, we deliberately selected geometrical features and electronic conditions that emulate high PF experimental results [15], and are expected to optimize electronic transport based on our previous theoretical works [16, 25, 26]. To facilitate the interpretation of the results, lattice thermal conductivities have been plotted in Fig. 4a (in green), alongside $S$ and $G$ (which are discussed later). Quantities in Fig. 4a have been normalized to the corresponding pristine geometry calculations. In Fig. 4 we have opted to plot the results against actual % porosity, $\varphi$, to show just how quickly thermal conductivity changes at low porosities. Note that $\varphi$ is defined as the ratio between the volume of the pores and the total simulation cell volume. The thermal conductivity calculations have been reported along with their 95% confidence interval as the error bars.

First, we investigate the effects of the SL barriers and/or pores on the thermoelectric properties of the 2D channels. We focus on acoustic phonon (AP) (*i.e.*, elastic) scattering conditions, where the results for $G$, $S$, and $PF$ are shown in Fig. 2 as functions of the number of pores (which vary in pairs of 2 for the NEGF geometries). For the electronic calculations, we have chosen degenerate conditions (with $E_F$ into the bands). These are beneficial to the PF — because high velocity electrons are utilized — and in that case the barrier heights, $V_B$, and the Fermi level, $E_F$, need to be placed at similar positions. As expected, introducing pores into an otherwise pristine system degrades the PF from its optimal value, and should be avoided if we only consider PF improvements. For $N = 0$, we notice that by raising $E_F$ high in the bands, the conductance of the SL channel (red dashed-dotted line in Fig. 2a) increases by 55% compared to the pristine channel (blue solid line). However, the Seebeck coefficient drops (see Fig. 2b), but overall the $PF$ increases in the SL channel by 23% compared to the pristine channel. This is an indication that the energy filtering, provided by potential barriers, is more effective at degenerate conditions as long as the energy of the current does not relax. However, the gradual introduction of pores in the region between the SL barriers causes significant degradation in the conductance and in the $PF$; namely, there is $14\% - 45\%$ reduction in the $PF$ from the SL reference channel depending on the number of pores $N$. Note that, for $N = 2$, the $PF$ is retained at a value higher than that of the pristine channel by 5%. It is also seen that the Seebeck coefficient of the SL channel slightly reduces as the number of pores increases (we discuss later the reason for this simultaneous decrease in $G$ and $S$). Also note that the behavior of the $PF$ of the SL structure follows a similar behavior to that of the pristine channel with pores and drops by $11\% - 45\%$ compared to a SL-only channel ($N = 0$).

The effect of the number of pores on the conductance is also reflected in the transmission versus energy. In Figs. 3a and 3b we show the transmission versus energy of a SL channel and of a pristine channel, respectively, for increasing number of pores. The transmission's energy dependence (for the pristine channel case with zero pores) is at first order linear, as it would also have been expected from the Boltzmann Transport Equation (BTE). Within the BTE, the transport distribution function (TDF) — which is proportional to the transmission [42], and whose integral in energy over the derivative of the Fermi distribution determines the conductivity — is given by $\Xi(E) = \tau v^2 g$, where $\tau$ is the relaxation time, $v$ is the bandstructure velocity, and $g$ is the density-of-states. Since the scattering rate is inversely proportional to the density-of-states (DOS), i.e. $\tau \sim 1/g$, then $\Xi(E) \sim v^2 \sim E$. Importantly, this behavior appears irrespective of dimensionality (even here the simulations are performed for low-dimensional channels). The TDF is thus at first order linear with energy, smearing out all sharp features of the DOS arising from the nature low-dimensional bands, which could have resulted in high Seebeck coefficients [39]. This linear dependence of the TDF and of the transmission across dimensions, further justifies our scaling of 3D measured data with 2D NEGF extracted coefficient ratios in our approach in Eq. 5, as no strong low-dimensional transmission features appear at room temperature, and even so, they would have been smeared out more when we take the ratios in Eq. 5.

We notice that as the number of pores increases from zero to four and then to ten the transmission is noticeably suppressed. This is in contrast to the case of nanoinclusions (NIs) [42] where minimal changes to the transmission are observed. We compare the transmissions for the two types of structures for a specific number of pores in Fig. 3c. Specifically, in Fig. 3c we show the transmission versus energy for a SL channel (red solid line) and for a pristine channel with four pores in each (blue dashed line). In the case of the SL, the transmission opens up as soon as the energy crosses the SL barrier height $V_{SL} = 0.05\text{eV}$ (around the SL Fermi level), but when this happens, the slope becomes larger compared to the channel without SL. However, at higher energies, the increase rate slows down until finally the transmission (red solid line) merges with that of the porous pristine channel (blue dashed line). This behavior has two consequences: i) The rising slope makes the SL transmission more robust in the presence of the pores (the three lines are basically overlapping around E = 0.05 in Fig. 3a). ii) The Seebeck coefficient, which is proportional to the slope of the transmission, drops slightly with increasing pore number, as observed in Fig. 2b (red line), matching the overall decrease in the slope of the lines in Fig. 3a as the pore number increases. This is in contrast with the Seebeck increase in the pristine case (blue line in Fig. 2b) which happens because the effective transmission shifts slightly to more positive values, when pores are included, as seen in Fig. 3b (green and magenta dashed lines) compared to the pristine geometry without pores (blue line).

The thermal conductivity of silicon is known to decrease rapidly at low porosity [45], as can be observed for the pristine channel in which pores are introduced (solid green line in Fig. 4a). At ~0.9 % porosity, an otherwise pristine geometry yields a sharp decrease in $\kappa_l$ of ~3 × the original value (*i.e.*, from 200.2 W m−1 K−1 to 65.3 W m−1 K−1). As porosity continues to increase, absolute changes in thermal conductivity are less marked, but nevertheless significant. For instance, between ~1 % and ~2 % porosities (corresponding to the 1 and 2 pore MD geometries, respectively) the thermal conductivity goes from 65.3 W m−1 K−1 to 43.6 W m−1 K−1, which corresponds to a 33% reduction in thermal conductivity. If not for the simultaneous degradation of the PF, this small change in porosity would correspond to a 33% improvement in $ZT$. At the highest porosity considered, which is a mere 4.6%, the $\kappa_l$ is already ~ 7.3 × lower than the pristine single-crystalline system.

For optimal doping conditions, superlattices could not only prove beneficial for the PF, as already discussed, they are also valuable in reducing the lattice thermal conductivity. In fact, for the geometries investigated herein, introducing grain boundaries, even when the material within the boundaries remains single-crystalline, yielded an impressive ~ 6.4 × reduction in $\kappa_l$. Note that some amorphicity exists near the boundaries where the different orientation crystals meet (see Fig. 1b). Compared to the lowest thermal conductivity geometry (*i.e.*, the higher porosity SL) the pristine system has a thermal conductivity that is ~15.8 × greater.

Finally, by using our results of $G$, $S$, $K_e$, and $\kappa_l$ to compute fractional changes in the values obtained from the literature, as indicated in Eq. 5, we can compute the effect of porosity and/or SLs on $ZT$. It is evident from Fig. 4 that the large $\kappa_l$ reduction coupled with the improvement of the *PF* that originates from the increase in conductance ($G$) under elastic scattering conditions is highly beneficial for the $ZT$ of the SL-only geometry. Merely by introducing grain boundaries, $ZT$ can be increased by approximately 7.5 × the value of single-crystalline Si. The additional introduction of pores brings $ZT$ to over 10 × the pristine system, as a result of maximum phonon scattering and enhanced electronic conductivity.

We mention here that in our NEGF simulations, we have considered only the effect of electron-acoustic phonon scattering. In practice, the positioning of the Fermi level is primarily achieved by doping, which introduces ionized impurity scattering (IIS). To estimate the effect of this on our conclusions, we performed Boltzmann Transport simulations calibrated to measured p-type Si data as in Refs. [16, 46, 47], and we compared the ratios of the conductivities for the phonon-limited and the phonon-plus-IIS cases, at the $E_F = Ec = 0$ eV, and $E_F = 0.05$ eV situations. We found that despite the large differences in the conductivities, the relative change of each conductivity as the $E_F$ changes is very similar, which indicates that the conclusions formed by the comparative methodology in Eq. 5 will still be the same if IIS where included.

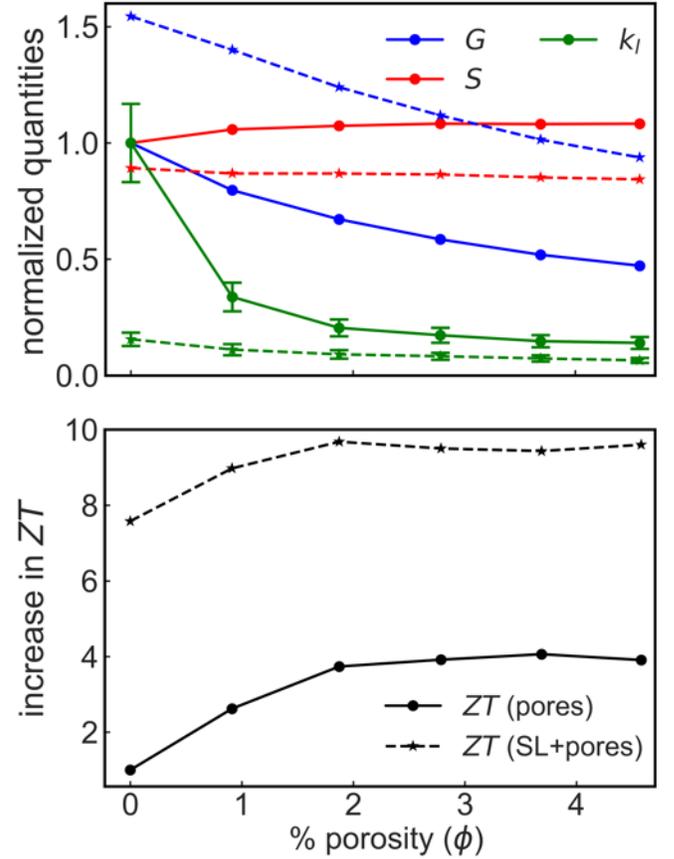

Fig. 4. (a) *S*, *G*, and $\kappa_l$ as a function of porosity and normalized against the same quantities for the pristine geometry, for the pores-only case (solid lines) and the SL + pores case (dashed lines). (b) Increase in ZT as a function of porosity compared to the pristine geometry, for the pores-only case (dashed line) and the SL+pores (solid line).

We can further determine from our results that beyond $\varphi \approx 2\%$, introducing additional porosity yields no gains. *I.e.*, for the geometries studied, beyond a ~ 2% porosity the reduction in the PF due to the increase in porosity overwhelms the simultaneous reduction in the thermal conductivity. This plateauing effect with porosity is observed in both geometries with and without SLs. While additional research is needed to generalize these results to other SL geometries and boundaries, we surmise these results could be useful in the design of both nanostructured channels and bulk or thin film nanoporous Si for thermoelectric applications.

## IV. CONCLUSIONS

In conclusion, we investigated the effects of embedded SL barriers and pores (for $\varphi < 5\%$) on the thermoelectric figure of merit in hierarchically nanostructured geometries, using equilibrium classical molecular dynamics simulations combined with the fully quantum mechanical non-equilibrium Green's function method. We show that upon nanostructuring to reduce $\kappa_l$, for the $PF$ it is beneficial to place the Fermi level high into the bands, *i.e.*, under highly degenerate conditions. In this way, utilizing the higher conducting states, benefits in the PF can be achieved, or at least its degradation is less severe. Nevertheless, we find that while the presence of SL barriers in combination with elevated Fermi levels, suppress the degrading influence of pores on the PF, the introduction of porosity to lower the thermal conductivity is only effective up to a point. In other words, we determine that an optimal porosity, situated at ~ 2% for our simulated geometries, exists beyond which point the degradation of the PF outweighs the degradation of the thermal conductivity. We believe that our findings can be of use to the on-going experimental efforts towards the design of advanced hierarchically nanostructured thermoelectric materials.

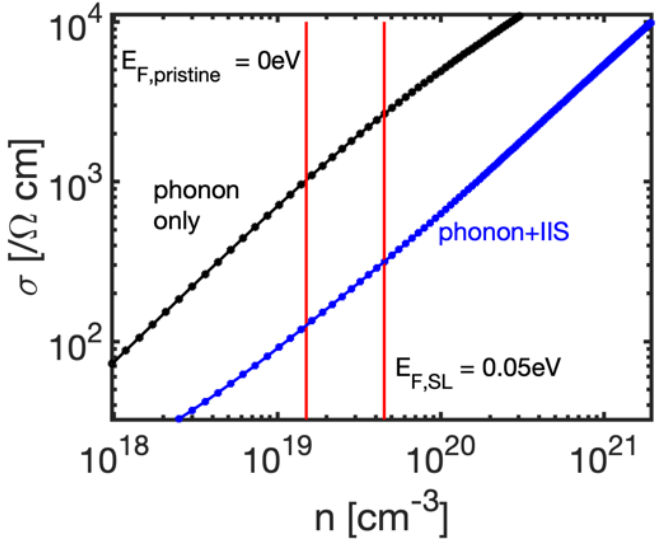

Fig. 1A. The conductivity of bulk p-type Si in the phonon-limited scattering case (black line) and the phonon plus ionized impurity scattering case (blue line). The vertical lines indicate the Fermi level $E_F$ position we use for the pristine and the superlattice structure we consider (but the simulations are for the bulk material). The slope of the two lines are the same, indicating similar changes for the two scattering cases at the two Fermi levels, of factor of 2.6 and 2.4, respectively.

## APPENDIX

The input for the $PF$ in Eq. 5 takes in the ratio of the disordered over the pristine conductances, $G_{features}/G_{pristine}$. In NEGF, we only consider electron-phonon scattering. In practice, high doping conditions are needed to achieve the elevated Fermi level at $E_F = V_{SL} = 0.05$ eV, which would introduce strong ionizing impurity scattering (IIS) as well. Thus, we examine here how those ratios, $G_{E_F = 0.05eV}/G_{E_F = 0eV}$, would be altered in the case where only phonon scattering is considered versus phonon + IIS. To investigate this, we have performed Boltzmann transport simulations (which are computationally less intensive), for phonon scattering only, and phonon plus ionized impurity scattering, and compare the ratios of the conductances for the two Fermi level cases. (We have matched our BTE simulator to the measured mobility of p-type Si as detailed in our previous work [16, 46, 47].) The conductivity results are shown in Fig. 1A, and with the vertical lines we indicate the position of the $E_F$ in the two cases. The ratio of the conductivities for each scattering case, at the points where the red lines cross each of the blue and black lines, are ~2.6 and 2.4, for the phonon only (black line), and phonon plus IIS cases (blue line). Thus, although we have not considered IIS in our NEGF simulations, since the ratio of the conductivities that enters into Eq. 5 would have been similar to the phonon-limited scattering case, our conclusions in the main text would not be altered.

## REFERENCES

1. Biswas, K., et al., *High-performance bulk thermoelectrics with all-scale hierarchical architectures.* Nature, 2012. **489**(7416): p. 414.
2. Mei, S., et al. *Boundaries, interfaces, point defects, and strain as impediments to thermal transport in nanostructures.* in *Reliability Physics Symposium (IRPS), 2017 IEEE International.* 2017. IEEE.
3. Gang, C., *Nanoscale heat transfer and nanostructured thermoelectrics.* IEEE Transactions on Components and Packaging Technologies, 2006. **29**(2): p. 238-246.
4. Liu, W., et al., *Modeling and Data for Thermal Conductivity of Ultrathin Single-Crystal SOI Layers at High Temperature.* IEEE Transactions on Electron Devices, 2006. **53**(8): p. 1868-1876.
5. Stranz, A., et al., *Thermoelectric properties of high-doped silicon from room temperature to 900 K.* Journal of electronic materials, 2013. **42**(7): p. 2381-2387.
6. Yang, R., G. Chen, and M.S. Dresselhaus, *Thermal Conductivity Modeling of Core–Shell and Tubular Nanowires.* Nano Letters, 2005. **5**(6): p. 1111-1115.
7. Boukai, A.I., et al., *Silicon nanowires as efficient thermoelectric materials,* in *Materials For Sustainable Energy: A Collection of Peer-Reviewed Research and Review Articles from Nature Publishing Group.* 2011, World Scientific. p. 116-119.
8. Yan, Z., et al., *Characterization of heat transfer along a silicon nanowire using thermoreflectance technique.* IEEE Transactions on Nanotechnology, 2006. **5**(1): p. 67-74.
9. Xu, B., et al., *Thermoelectric Performance of $Si_{68}Ge_{02}$ Nanowire Arrays.* IEEE Transactions on Electron Devices, 2012. **59**(12): p. 3193-3198.
10. Choday, S.H., M.S. Lundstrom, and K. Roy, *Prospects of Thin-Film Thermoelectric Devices for Hot-Spot Cooling and On-Chip Energy Harvesting.* IEEE Transactions on Components, Packaging and Manufacturing Technology, 2013. **3**(12): p. 2059-2067.
11. *Nanostructure design for drastic reduction of thermal conductivity while preserving high electrical conductivity AU - Nakamura, Yoshiaki.* Science and Technology of Advanced Materials, 2018. **19**(1): p. 31-43.
12. Joshi, G., et al., *Enhanced thermoelectric figure-of-merit in nanostructured p-type silicon germanium bulk alloys.* Nano letters, 2008. **8**(12): p. 4670-4674.
13. Thumfart, L., et al., *Thermal transport through Ge-rich Ge/Si superlattices grown on Ge (0 0 1).* Journal of Physics D: Applied Physics, 2017. **51**(1): p. 014001.
14. Weber, L. and E. Gmelin, *Transport properties of silicon.* Applied Physics A, 1991. **53**(2): p. 136-140.

15. Lorenzi, B., et al., *Paradoxical enhancement of the power factor of polycrystalline silicon as a result of the formation of nanovoids.* Journal of electronic materials, 2014. **43**(10): p. 3812-3816.

16. Neophytou, N., et al., *Simultaneous increase in electrical conductivity and Seebeck coefficient in highly boron-doped nanocrystalline Si.* Nanotechnology, 2013. **24**(20): p. 205402.

17. Bennett, N.S., et al., *Dislocation loops as a mechanism for thermoelectric power factor enhancement in silicon nano-layers.* Applied Physics Letters, 2016. **109**(17): p. 173905.

18. Neophytou, N., et al., *Power factor enhancement by inhomogeneous distribution of dopants in two-phase nanocrystalline systems.* 2014. **43**(6): p. 1896-1904.

19. Bulusu, A. and D.G. Walker, *Quantum Modeling of Thermoelectric Properties of Si/Ge/Si Superlattices.* IEEE Transactions on Electron Devices, 2008. **55**(1): p. 423-429.

20. Thesberg, M., et al., *The fragility of thermoelectric power factor in cross-plane superlattices in the presence of nonidealities: a quantum transport simulation approach.* Journal of Electronic Materials, 2016. **45**(3): p. 1584-1588.

21. Kearney, B., et al., *From amorphous to nanocrystalline: the effect of nanograins in an amorphous matrix on the thermal conductivity of hot-wire chemical-vapor deposited silicon films.* Journal of Physics: Condensed Matter, 2018. **30**(8): p. 085301.

22. Foster, S., M. Thesberg, and N. Neophytou, *Thermoelectric power factor of nanocomposite materials from two-dimensional quantum transport simulations.* Physical Review B, 2017. **96**(19): p. 195425.

23. Ahmad, S., et al., *Boosting thermoelectric performance of p-type SiGe alloys through in-situ metallic YSi2 nanoinclusions.* Nano Energy, 2016. **27**: p. 282-297.

24. Liu, M. and X. Qin, *Enhanced thermoelectric performance through energy-filtering effects in nanocomposites dispersed with metallic particles.* Applied Physics Letters, 2012. **101**(13): p. 132103.

25. Kim, R. and M.S. Lundstrom, *Computational study of the Seebeck coefficient of one-dimensional composite nano-structures.* Journal of Applied Physics, 2011. **110**(3): p. 034511.

26. Neophytou, N. and H. Kosina, *Optimizing thermoelectric power factor by means of a potential barrier.* Journal of Applied Physics, 2013. **114**(4): p. 044315.

27. Green, M.S., *Markoff random processes and the statistical mechanics of time-dependent phenomena. II. Irreversible processes in fluids.* The Journal of Chemical Physics, 1954. **22**(3): p. 398-413.

28. Kubo, R., *Statistical-mechanical theory of irreversible processes. I. General theory and simple applications to magnetic and conduction problems.* Journal of the Physical Society of Japan, 1957. **12**(6): p. 570-586.

29. Stillinger, F.H. and T.A. Weber, *Computer simulation of local order in condensed phases of silicon.* Physical review B, 1985. **31**(8): p. 5262.

30. Lee, Y. and G.S. Hwang, *Force-matching-based parameterization of the Stillinger-Weber potential for thermal conduction in silicon.* Physical Review B, 2012. **85**(12): p. 125204.

31. Howell, P., *Comparison of molecular dynamics methods and interatomic potentials for calculating the thermal conductivity of silicon.* The Journal of chemical physics, 2012. **137**(22): p. 2129.

32. Broughton, J. and X. Li, *Phase diagram of silicon by molecular dynamics.* Physical Review B, 1987. **35**(17): p. 9120.

33. Coquil, T., J. Fang, and L. Pilon, *Molecular dynamics study of the thermal conductivity of amorphous nanoporous silica.* International Journal of Heat and Mass Transfer, 2011. **54**(21-22): p. 4540-4548.

34. Schelling, P.K., S.R. Phillpot, and P. Keblinski, *Comparison of atomic-level simulation methods for computing thermal conductivity.* Physical Review B, 2002. **65**(14): p. 144306.

35. de Sousa Oliveira, L. and P.A. Greaney, *Method to manage integration error in the Green-Kubo method.* Physical Review E, 2017. **95**(2): p. 023308.

36. Plimpton, S., *Fast parallel algorithms for short-range molecular dynamics.* Journal of computational physics, 1995. **117**(1): p. 1-19.

37. Datta, S., *Electronic transport in mesoscopic systems.* 1997: Cambridge university press.

38. Vargiamidis, V., S. Foster, and N. Neophytou, *Thermoelectric power factor in nanostructured materials with randomized nanoinclusions.* physica status solidi (a), 2018: p. 1700997.

39. Neophytou, N. and H. Kosina, *Atomistic simulations of low-field mobility in Si nanowires: Influence of confinement and orientation.* Physical Review B, 2011. **84**(8): p. 085313.

40. Qiu, B., et al., *First-principles simulation of electron mean-free-path spectra and thermoelectric properties in silicon.* EPL (Europhysics Letters), 2015. **109**(5): p. 57006.

41. Persson, M.P., et al., *Orientational dependence of charge transport in disordered silicon nanowires.* Nano letters, 2008. **8**(12): p. 4146-4150.

42. Vargiamidis, V. and N. Neophytou, *Hierarchical nanostructuring approaches for thermoelectric materials with high power factors.* Physical Review B, 2019. **99**(4): p. 045405.

43. Lundstrom, M., *Fundamentals of carrier transport (Cambridge Univ Pr, 2000).* Cited on: p. 45.

44. Neophytou, N. and M. Thesberg, *Modulation doping and energy filtering as effective ways to improve the thermoelectric power factor.* Journal of Computational Electronics, 2016. **15**(1): p. 16-26.

45. Verdier, M., K. Termentzidis, and D. Lacroix, *Crystalline-amorphous silicon nano-composites: Nano-pores and nano-inclusions impact on the thermal conductivity.* Journal of Applied Physics, 2016. **119**(17): p. 175104.

46. Jacoboni, C. and L. Reggiani, *The Monte Carlo method for the solution of charge transport in semiconductors with applications to covalent materials.* Reviews of modern Physics, 1983. **55**(3): p. 645.

47. Masetti, G., M. Severi, and S. Solmi, *Modeling of carrier mobility against carrier concentration in arsenic-, phosphorus-, and boron-doped silicon.* IEEE Transactions on electron devices, 1983. **30**(7): p. 764-769.

48. Neophytou, N. and H. Kosina, *Effects of confinement and orientation on the thermoelectric power factor of silicon nanowires.* Physical Review B, 2011. **83**(24): p. 245305.

49. Neophytou, N., *Prospects of low-dimensional and nanostructured silicon-based thermoelectric materials: findings from theory and simulation.* The European Physical Journal B, 2015. **88**(4): p. 86.